# Designing the W3C Open Annotation Data Model

**Robert Sanderson**
Los Alamos National Laboratory
Los Alamos
NM 87544, USA
rsanderson@lanl.gov

**Paolo Ciccarese**
Massachusetts General Hospital
Boston
MA 02114, USA
paolo.ciccarese@gmail.com

**Herbert Van de Sompel**
Los Alamos National Laboratory
Los Alamos
NM 87544, USA
herbertv@lanl.gov

**ABSTRACT**
The Open Annotation Core Data Model specifies an interoperable framework for creating associations between related resources, called annotations, using a methodology that conforms to the Architecture of the World Wide Web. Open Annotations can easily be shared between platforms, with sufficient richness of expression to satisfy complex requirements while remaining simple enough to also allow for the most common use cases, such as attaching a piece of text to a single web resource. This paper presents the W3C Open Annotation Community Group specification and the rationale behind the scoping and technical decisions that were made. It also motivates interoperable Annotations via use cases, and provides a brief analysis of the advantages over previous specifications.

**ACM Classification Keywords**
H.5.4 [Information Interfaces and Presentation]:
Hypertext/Hypermedia – *Architectures, Navigation*.

## INTRODUCTION

Annotating, the act of creating associations between distinct pieces of information, is a pervasive activity on the web, but currently lacks a uniform approach. The interactive and collaborative "Web 2.0", where Web citizens are part of a global community rather than mere consumers of information, has produced an environment that both socially encourages and engenders technical challenges for annotating. Currently, comments are made about online resources using tools built in to the hosting web site, external web services, or the functionality of non-browser annotation clients. In addition to commenting on resources in a wide variety of social media platforms, there are a plethora of closed and proprietary web-based "sticky note" systems and stand-alone multimedia annotation systems. The primary complaint about these types of systems is that the user-created annotations cannot be shared or reused due to a deliberate "lock-in" strategy within the environments where they were created. The minimum requirement for any solution in this domain is a common approach to modeling and expressing annotations, and this is the primary aim of the Open Annotation work.

The Open Annotation data model provides an extensible, interoperable framework for expressing annotations such that they can easily be shared between platforms, with sufficient richness of expression to satisfy complex requirements while remaining simple enough to also allow for the most common use cases, such as attaching a piece of text to a single web resource. The intended use of this interoperability is either for sharing public annotations with others or for the migration of private annotations between systems or devices. Therefore, the annotations must be able to be integrated into existing collections and reused without loss of significant information.

This paper introduces the technical model created by the W3C Open Annotation Community Group, of which the authors are co-chairs and editors, but focuses on the reasoning behind the decisions that were made throughout the process in order to provide additional information that may not be apparent from the specification itself [15].

**Aims and Values for the Open Annotation Data Model**

Before turning to the data model and the decisions that were made, it is important to understand the aims and values that guided the decision making process.

The primary aim of the work is to create a standard description mechanism for sharing annotations between systems. This, in turn, leads to several fundamental value judgments regarding how the model should be designed. The five values, in order of decreasing importance are:

1. A single, consistent model is the highest priority. Interoperability is greatly decreased if there are either separate competing models, or a single model with many different ways to address the same use case. In these situations consuming applications must implement all of the possibilities to remain compatible, which is similar to the lack of any standard at all. The same model should accommodate all of the required use cases that are considered to be within the scope of the effort.

2. Re-use of existing ontologies and standards, without modification and where appropriate, is important to ensure interoperability. The Web Architecture [8] is considered the baseline on top of which RDF and recognized ontologies are added as an interoperability infrastructure.

3. Implementation costs for both producers and consumers should be kept to a minimum to encourage widespread adoption. This includes both the number of URIs that are required to be maintained as well as the number of lines of code required for implementations.

4. The model is abstract and must not require specific storage or internal modeling requirements. Interoperability is between systems, the technologies used to maintain the annotation data within a system are not to be specified or assumed.

5. The number of triples, or the number of bytes when those triples are serialized, to be transferred between systems should be kept as low as possible without compromising the previous values.

With those values in mind, we turn to some of the use cases that drive the design of the data model.

**USE CASES**

While there are far too many use cases to discuss them all in detail, it is useful to consider the breadth of scenarios that are considered in scope for the model and serve as input to the design process.

The most common use case for online annotation is simply attaching a textual comment to an existing web page. This could take the form of an inline threaded forum or commenting system, an overlaid digital sticky-note, or a sidebar application in the browser. Attaching the annotation to a particular segment of the page is also a very common requirement.

An extension to this case is when the comment or the resource being annotated is not textual. It could be a video or audio file, for example, which is about an image or data set. Video comments are becoming increasingly common with the ability to easily upload from mobile devices or computers with integrated webcams. The ability to annotate arbitrary resources is particularly important in scholarly use cases, where art, performances, large scale high energy physics datasets, medical or astronomical images and many others are the topic of discussions.

Tagging resources, either with a simple string or a semantic URI, is also a very common case. As applications treat tags and comments very differently, these two situations must be carefully distinguished. The URI used as a semantic tag should not be dereferenced, when rendering the annotation, whereas the URI of a video should be. If the "like" feature of current social media sites were implemented as a tag annotation, it alone would dwarf all other use cases combined.

Machines also create annotations to be consumed by other machines. Text Mining applications can quickly annotate every word of a document with its part of speech, and then build upon those annotations to create phrases, grammar trees, and identify entities plus the relationships between them. This is particularly common in biomedical research, but also for annotation and processing of news stories.

There are annotations that do not have an explicit comment. Instead they are pointers to a resource like a bookmark, or highlight a section of a resource. The highlight's stylistic features may convey additional information, such as a red strike-through of text often means that it should be deleted.

Scholarly commentary and discussion provides many use cases and examples where existing models are insufficient to capture the richness of the possible semantics. Even the simplest case of comparing or contrasting two or more resources is not possible in systems where an annotation must be about exactly one resource. These use cases can easily be recast into requirements of normal web citizens: comparing the similarities between two editions of a novel is the same, in modeling terms, as comparing the terms and conditions documents of different services or platforms.

Finally, editing requests expressed as annotations may provide both information as to why a change should be made at the same time as providing replacement text or other content. Such annotations could be seen as micro-edits within wiki-like systems, or annotations on documents that are being collaboratively created.

After explaining the data model in detail, an example use case of annotating the W3C Open Annotation Community Group home page[1] will be given.

**PREVIOUS WORK**

During the process of designing the Open Annotation data model, more than 50 existing systems and ontologies were researched[2] and mappings generated where possible to ensure that the model covered as many of the existing tools as possible. It is not possible to describe them all in detail, however exemplary work has been done previously to summarize and categorize many of the systems available.

Marshall's papers [11,12] on annotation provide an excellent introduction to the topic, and her insight into use cases and individuals' uses of annotation in invaluable. Agosti has also published prolifically [2,3] about annotation, including formalizing a service-centric, rather than resource-centric, framework. A comprehensive requirements checklist can easily be derived from the works of these two authors alone.

More specific summaries and reviews of the landscape have been performed recently by Haslhofer et al. [7] and by Adriano and Ricarte in their article "Essential Requirements for Digital Annotation Systems" [1]. Further analysis is always required, as the domain is rapidly improving along with technologies and our understanding of the web.

---

[1] http://www.w3.org/community/openannotation/

[2] http://www.openannotation.org/wiki/index.php/Main_Page

The three systems that provided the most significant input to the model are described and compared to further the understanding of the background in which the Open Annotation data model was produced.

The first well-known RDF based Annotation system was Annotea [9]. It had a minimal data model, with only one class and the seven properties listed in Table 1. An extensive protocol recommendation accompanied it as to how to transfer annotations between client and a dedicated server.

| a:annotates | oa:hasTarget |
| --- | --- |
| a:author | oa:annotatedBy |
| a:body | oa:hasBody (embedded) |
| a:context | *(Specific Resources)* |
| a:created | oa:annotatedAt |
| a:modified | oa:annotatedAt |
| a:related | oa:hasBody (external) |

**Table 1. Annotea / Open Annotation mapping**

Annotea predates the resource-centric perspective made explicit in the Architecture of the World Wide Web, and hence lacks the distinction between the Annotation document and Annotation concept. The context node was very poorly defined, and could only apply to a single target. As expected, the understanding of annotation has come a long way in a dozen years.

Annotea was extended several times, including by Hunter [16] to allow multiple annotation targets for example. The Vannotea system also allows for multiple media types including video annotation and SVG described media segments. Secondly, Haslhofer's LEMO [7] system extended Annotea to permit complex segment descriptions and multimedia annotations. Along with many other systems, these extensions show significant dissatisfaction with Annotea as a framework.

The Open Annotation Collaboration [15] was one of the direct precursors to the Open Annotation Community Group. The beta model introduced many of the concepts that were subsequently adopted by the working group, albeit in different and improved forms. Specific Resources were expressed as constraints, and not divided into State, Selector, Style and Scope. There was significant ambiguity about the interpretation of multiple targets, and multiple bodies were not allowed at all. Instead of the richer motivations, subclasses of Annotation were used which could have led to significant fractures between communities. Semantic annotation and tagging were not addressed at all.

The Annotation Ontology (AO) [5] was the other model used as starting point for the Open Annotation Community Group. The model also made use of multiple bodies, provided solutions for semantic annotation and tagging and provided an initial implementation of motivations. However, AO did not provide clear guidelines on how to organize multiple bodies and the semantic tags were supported through a separate relationship (`ao:hasTopic`), rather than `oa:hasBody`, to reflect the different nature of the aboutness. AO did not provide the means to express Specifiers such as State, Style and Scope, which had to be addressed by the implementers using existing or local solutions.

**OPEN ANNOTATION DATA MODEL**

In general terms, an Open Annotation expresses the relationship between two or more resources by means of an RDF graph. The three basic resources described in the graph are the Annotation itself, which identifies the concept of this relationship, the Body and the Target. The Body is a resource such as a comment or other content that is normally somehow "about" the Target. This "aboutness" may be further clarified, and extends to notions including identifying, classifying, describing, and moderating.

The Body and Target resources, as depicted in Figure 1, may be of any media type: a video comment is as valid as a textual one for the purposes of an Annotation. The general content type (Text, Audio, Video, etc.) is conveyed as a class, and the specific media type, if known, using the `dc:format` property.

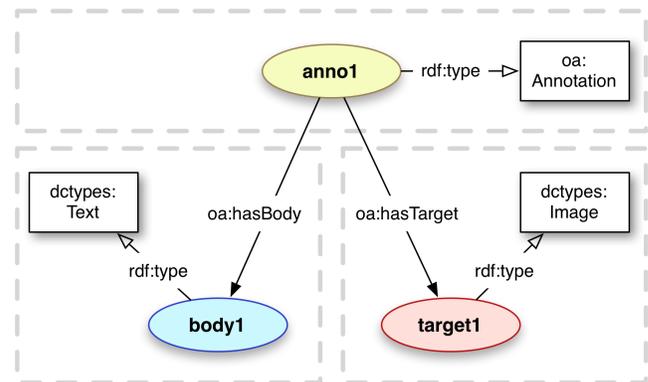

**Figure 1. Basic Annotation Data Model**

The `oa:hasBody` and `oa:hasTarget` relationships, used to connect the Annotation and the Body and Target resources, can be repeated. This allows a single comment to be about multiple resources, for multiple tags to be applied at the same time to a resource, and combinations thereof. This was not possible in previous systems. The semantics of the concept expressed by the Annotation are that each of the Bodies is individually related to each of the Targets.

**Embedded Resources**

Many previous and existing annotation systems have used a property with a string literal to represent textual Bodies. However given that the highest priority is for a single consistent model, the Body should always be a resource rather than a literal. Following the second priority for re-use, the "Content in RDF" specification [10] was discovered and recommended for use. The specification makes use of a class `cnt:ContentAsText` which represents the content, and includes it directly in the graph using a property `cnt:chars`.

This node should be able to be referenced externally, for example to point out a typo, and thus a UUID URN is recommended. On the other hand, there are many situations when minting and maintaining this URI is an unnecessary burden, in which case an RDF blank node may be used. The most recent RDF specification also recommends a skolemization technique to replace blank nodes with IRIs [6] that may be adopted. This is depicted in Figure 2 with a UUID for the Body resource.

As a blank node does not have an identifier, there is only one additionally required triple (the triple with `cnt:chars` as predicate) in the blank node case when compared to an approach that would make use of a literal string, and thus the cost in priority five is also minimal. Other arguments in favor for introducing a resource even for simple textual bodies include the inability to have a single punning property with a range of either a literal or a resource in OWL-DL, the consistency of the recommended JSON-LD serialization, the ability to attach additional metadata such as media type, class, authorship and so forth.

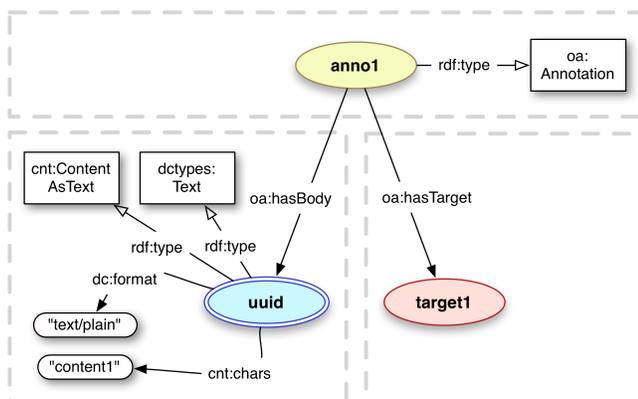

**Figure 2. Embedded Textual Body**

**Motivations**

The "aboutness" that is described in the Annotation is conveyed using an instance of the `oa:Motivation` class. This is a subclass of `skos:Concept`, and conveys a reason for the creation of the Annotation. Previous systems have used subclassing of the Annotation class for this purpose, but the use of the SKOS vocabulary [13] allows for Motivations to be inter-related between communities with more meaningful distinctions than a simple class hierarchy. Multiple motivations from different communities can be used, without inheriting semantics associated with classes such as the cardinality, domain or range of properties. The specification provides further guidance on how best to extend the basic list of Motivations provided.

This maintains a single model, albeit slightly different to previous systems, and reuses the existing SKOS standard, while keeping the costs in terms of triples and URIs to maintain low. Figure 3 depicts an Annotation that is motivated by editing the target resource.

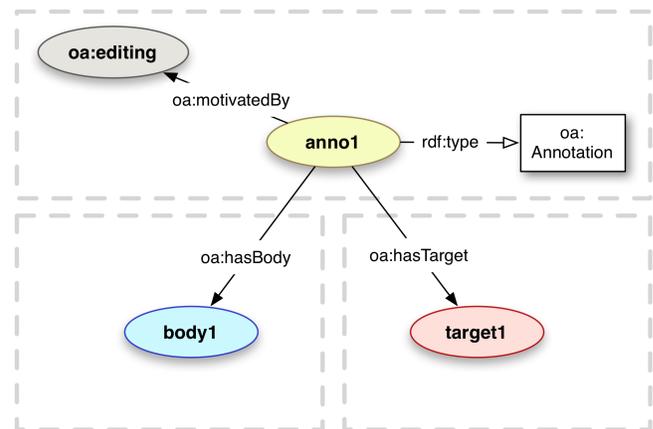

**Figure 3. Motivations**

**Tagging Requirements**

Tagging a resource with a short textual string and semantic tagging with a URI are common annotation use cases. Tags are often keywords, labels or very short descriptive phrases used for organization, description, or discovery of the resource being tagged. The "aboutness" of tags is therefore quite dissimilar to the typical commentary Annotations, and the Motivations described in the previous sections can be used to express this. As language is inherently ambiguous, tagging a resource with a URI is also common in the semantic web to avoid the polysemy of, for example, the bank of a river and a financial institution.

Consuming applications will often treat tags in a very different way to commentary or other descriptive annotations. Instead of presenting every tag individually, they may be rendered in a tag cloud that gives the most prominence to the tags that have been frequently repeated, and thus the distinction between these two use cases must be clear. For semantic tags it is even more important, as the URI should not be dereferenced and the representation displayed. Instead it should be treated in the same way as the textual tag, but enables the client to distinguish between otherwise identical string literals.

Following the basic model, Body and Target resources are distinguished by means of classes. Open Annotation defines an `oa:Tag` class to distinguish between tags and non-tags. As textual tags are embedded within the graph, this class can also be applied alongside the `cnt:ContentAsText` class to a resource identified by a URI or to a blank node. In this way, there is a single, consistent model for tags, semantic tags, embedded resources and regular dereferencable resources, fulfilling the first priority while enabling many use cases to be clearly modeled.

The oa:tagging Motivation should also be used to signal the reason for the creation of the Annotation, depicted in Figure 4. This is insufficient by itself as a distinguishing feature, as a single Annotation may associate both a regular comment and a tag with the Target, and a consuming application would be unable to tell which role each took.

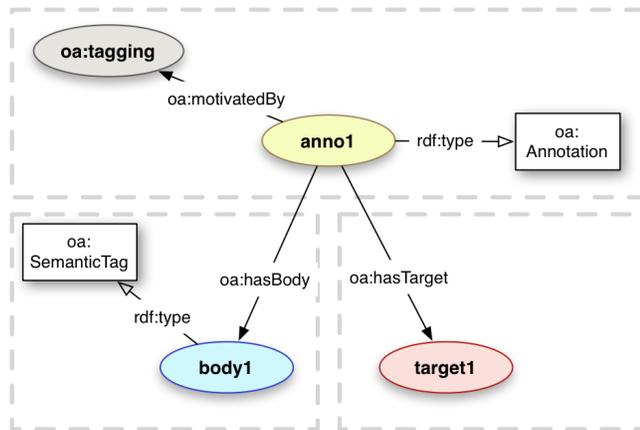

Figure 4. Semantic Tagging

**Fragment URIs**

Many Annotations are about, or otherwise related to, part of the Target resource rather than its entirety. Resources can be arbitrarily large, and annotations arbitrarily precise as to their segment of interest. In the Architecture of the World Wide Web [8], segments of resources are identified using URIs with a Fragment component, henceforth "Fragment URIs". These URIs at the same time identify the segment of interest and describe how to extract it from the resource.

However, there are several restrictions about the use of Fragment URIs that make them incapable of expressing all of the information necessary for the annotation use cases. In particular:

- Most fragments are defined with respect to individual media types, and not every media type has a fragment specification. Thus if Fragment URIs were the only means of selecting the segment of interest, only (segments of) resources of those media types could be annotated. This would severely limit the usefulness of the specification and hamper its adoption.

- Even if a media type does have a fragment definition, it is often not possible to describe the segment of interest in sufficient detail. For example, HTML fragments cannot be used to describe an arbitrary span of text, and W3C Media Fragments [17] cannot describe circular regions.

- It is not possible to accurately determine what a particular Fragment URI identifies without its media type, as the same fragment string can be made possible by different competing specifications. For example, the same fragment could identify either a semantic resource in RDFa, or a section of the embedding HTML document.

This is a cause of conflict and concern for the design of the Open Annotation data model, as the priorities described pull the result in very different directions.

The first priority would mean not using fragments and instead designing a single system that was capable of fulfilling the requirements. However this contradicts all of the other priorities! Fragments are a fundamental part of the Web Architecture, and have been standardized in both the IETF and the W3C. Creating our own description would be in direct violation of the second priority.

The implementation costs, as shall be seen in the next section, of a design to fulfill the requirements is significantly greater than using an existing URI with a fragment component. As the majority of the use cases do not require this extra level of detail, it is a burden in terms of the number of lines of code and URIs to maintain.

The decision was made that although introducing multiple methods of describing segments of interest contravenes the first priority, the value of allowing Fragment URIs outweighs this. A further argument was that the Fragment URI case was not prevented anyway, as any URI may be used as the identifier in the Body or Target role of the Annotation, which necessarily includes Fragment URIs.

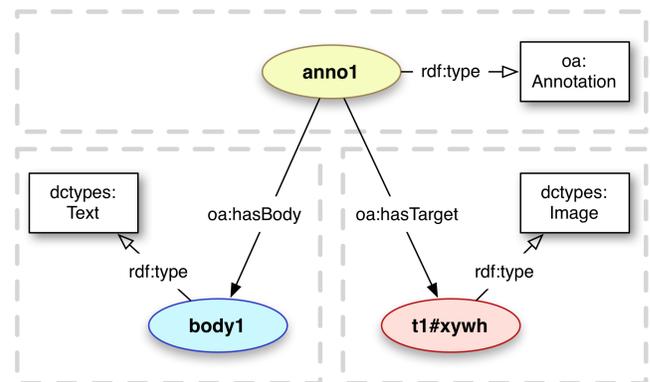

Figure 5. Fragment URI as Target

Figure 5 shows the same structure as Figure 1, but with the Target being a W3C Media Fragment URI that describes and identifies a rectangular section of the image.

## DATA MODEL MODULES

While the core of the Open Annotation Data Model is sufficient to handle a majority of annotation use cases, a significant number of cases remain that are not covered. These include segments of resources that cannot be described using Fragment URIs and especially text selection, applying styles to Annotations, annotating changing resources and resources that have multiple representations, and being precise about the semantics of the use of multiple resources. Most of these cases are supported by the Specific Resources Module, and the final one by the Multiplicity Constructs Module.

### Resource Segment Selection

In order to handle resource segments, the Open Annotation data model distinguishes between a resource that *identifies* the segment of interest and a resource that *describes* how to extract it from the representation of the full resource. The resource that identifies the segment is called the Specific Resource, the resource that describes how to extract the correct segment is a Selector, and the full resource is termed the Source Resource in this context. For example, a circular region of an image would be the Specific Resource, an SVG based Selector could be used to describe its location and size, and the full image is the Source.

Selectors, as in Figure 6, are intended to be reusable across Annotations, they are just resources after all, and hence are not attached directly to the Source. Doing so would also violate the global scope of assertions in RDF. The Specific Resource may also be re-used by other Annotations, so long as all of its properties hold true for the second and subsequent uses. The Specific Resource and Selector should have UUID URNs to enable this, but may be modeled as blank nodes to avoid the URI maintenance cost.

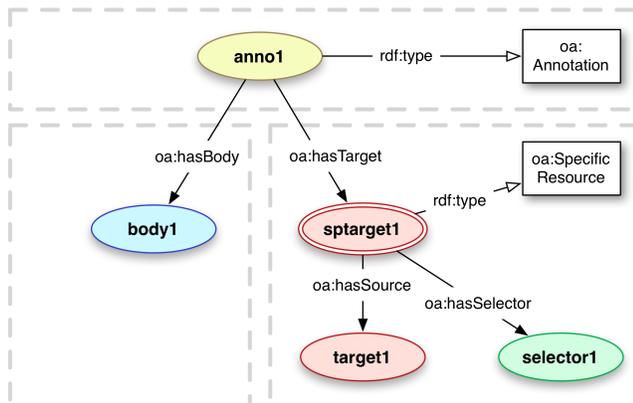

**Figure 6. Specific Resource and Selector**

### Fragment Selectors

As previously described, having two approaches to handle segments violates the single consistent model priority, however a transition mechanism to alleviate this is introduced in the form of the `oa:FragmentSelector` class.

This class of Selector takes the fragment component of the URI and records it within the Annotation's graph using the `rdf:value` property. This is the best of both worlds, as the specification that defines a fragment syntax is re-used at minimal cost to the implementer. There is a consistent model using oa:hasSource that conveys the URI of the full resource for querying, regardless of the method used to describe the segment. Furthermore, the fragment interpretation can be disambiguated by asserting, using the `dcterms:conformsTo` relationship, that the Selector uses a particular specification; e.g. the Fragment Selector in Figure 7 conforms to the Media Fragments specification.

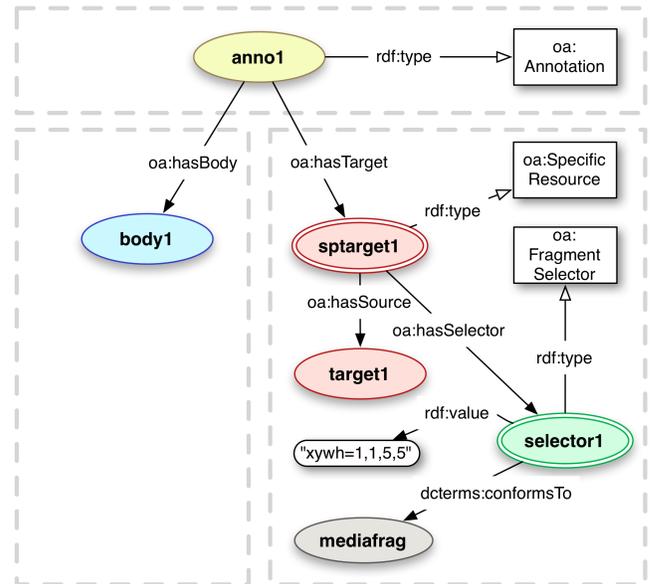

**Figure 7. Fragment Selector**

The complete Fragment URI may be easily reconstructed by taking the URI of the Source resource, plus "#", plus the literal from the Selector. For example in Figure 7, it would the URI `"target1#xywh=1,1,5,5"`. Publishing systems may also rewrite Fragment URIs using Fragment Selectors.

### Text Selectors

As the one existing fragment specification for selecting text ranges is only applicable for the text/plain media type, it is necessary to define our own method capable of handling resources in other textual formats. There are two equally valid use cases that make it impossible to define a single text selector class.

The first use case is a textual resource that remains static and the content of which should be protected. A classic example is an eBook or PDF document where the text cannot or should not be shared, but it is still desirable to annotate it. In this case, the position of the start and end of the text selection is recorded in an `oa:TextPositionSelector`. This prevents people from selecting the entire text, or even consecutive sections, and publishing Annotations that quote it verbatim, which would potentially violate Digital Rights Management restrictions. This model is depicted in Figure 8, below.

The second use case is when the content is more dynamic, such as a web page or document being edited and the position of the selection will very quickly become invalid as text blocks are added, moved and deleted. In this case, an `oa:TextQuoteSelector` is recommended. It quotes the selection and a range of characters immediately before and after to provide context in case the selection alone is not unique within the document.

Given the ubiquity of both of these scenarios, it was deemed impossible to have a single textual selector that covered both situations.

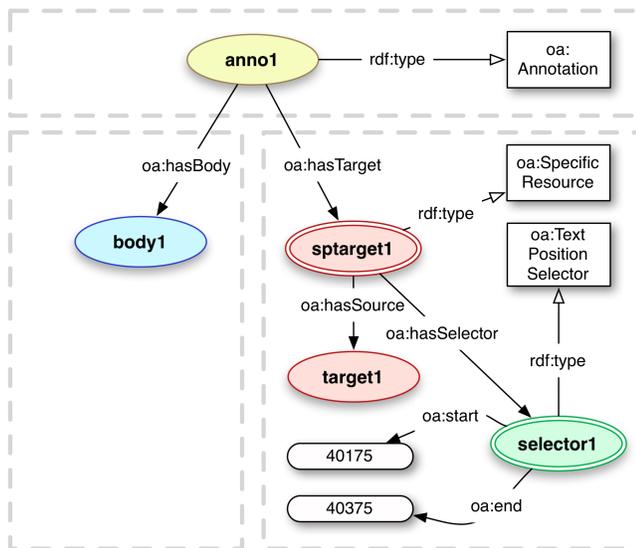

**Figure 8. Text Position Selector**

### The Dynamic Web

Especially given the widespread adoption of Web 2.0 techniques, the resources that users encounter and may want to annotate are increasingly dynamic. It is important to be able to address the state of resources that change over time or are available in multiple formats due to HTTP Content Negotiation or by being responsive to different devices such as tablets versus desktop computers. There is also, at the same time, more research and experimentation to be done regarding the meaningful identification and description of scripted resources. The Open Annotation data model recognizes both the immediate need and that there will be both challenges and solutions appearing in the future in this domain.

The essence of the problem can be reduced, in Web Architecture terms, to the distinction between annotating a resource and annotating a particular representation of that resource. As the agent (either software or human) creating the Annotation is using a particular representation, it is most frequently the latter that is intended. To resolve this issue, the data model introduces the `oa:State` class to record the information needed for future clients to reproduce the representation of the resource that was used.

There are currently two sub-classes of `oa:State` defined. The first is `oa:TimeState`, which conveys the time at which the resource was used or is appropriate for the Annotation. It does this either by simply recording the timestamp and leaving it up to the client to discover a representation, potentially using the Memento protocol [18], or by linking to an archived copy of the resource such as in the Internet Archive[3]. The use of this sub-class is depicted in Figure 9.

And secondly, `oa:HttpRequestState` conveys the headers that must be used to obtain the correct representation. This allows for specific HTTP Content Negotiation, as well as recording the user-agent or other necessary HTTP headers.

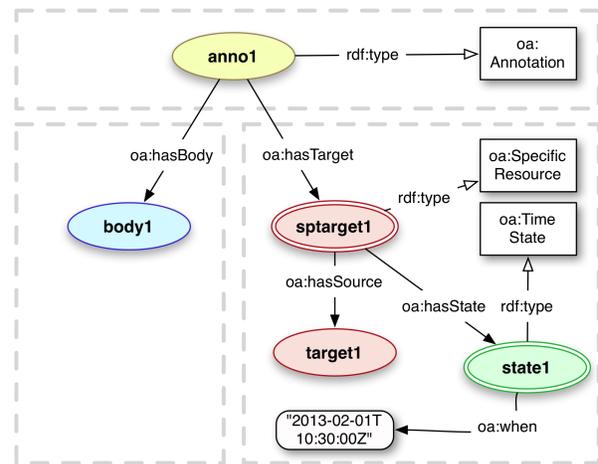

**Figure 9. Time State**

It is expected that other State sub-classes will be defined in the future to cover additional use cases, either within the Open Annotation Community Group, or by further communities.

### Highlighting and Styles

The interpretation by a user of a particular Annotation may rely on clients rendering it consistently. For example, if the comment refers to part of an image which the annotator highlighted in yellow, and compares it to a segment that she highlighted in red, then these colors need to be maintained for the text to make sense. Furthermore, several uses of annotation convey all of their information by stylistic features alone, from red-strike through of a text segment being the well-known method of suggesting a deletion, or personal conventions such as highlighting segments of text to refer back to in green, versus those disagreed with in red, for example. In these cases, there may be no Body resource at all.

---

[3] http://web.archive.org/

The Open Annotation data model reuses the CSS specification [4] to fulfill these requirements, which promotes the separation of the presentation layer and the content layer for HTML and XML documents. As an Annotation is neither of these, some additional features are required in the model to support it. The style resource itself is attached to the Annotation, as it may include style descriptions for more than one resource, with the `oa:styledBy` relationship. Only the class selector construction is used in the CSS document and the class it expresses is attached to a Specific Resource with the `oa:styleClass` property, as depicted in Figure 10. The CSS document may be reused across multiple Annotations, perhaps by an application that associates a default set of styles for a human user to select from. It may also be embedded using the Content in RDF specification in the same manner as embedding a textual body resource.

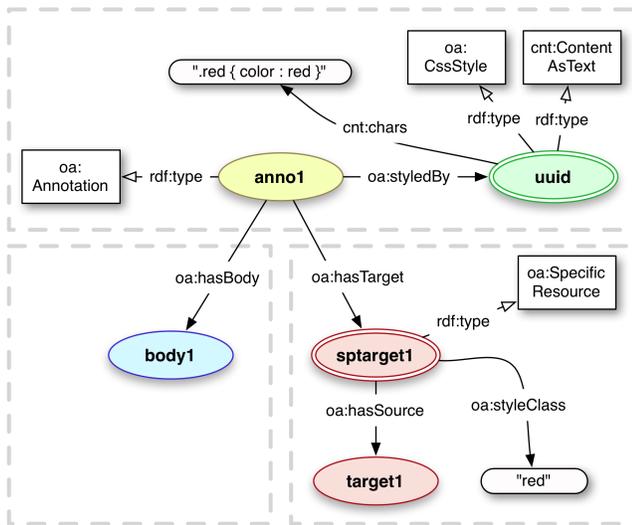

**Figure 10. Styling an Annotation**

A previous version of the specification had the CSS value block, e.g. `"{color:red}"` without the css selector, attached directly to the Specific Resource. This was changed in order to ensure that the style resource was a valid CSS document and to mirror the common usage with a single reference to a stylesheet and individual resources inheriting styles via class, identity or element name.

**Semantics of Multiplicity**

There are many use cases where an Annotation comprises multiple Bodies, Targets or both. The semantics for multiple occurrences of `oa:hasBody` and `oa:hasTarget` are that each Body individually is about each Target, but this is not always the case. It is useful to have several Body resources as alternatives, for example to have the same content in different languages or formats, of which only one should be selected and rendered as appropriate to the user. A second scenario, for Targets, would be when the comment is comparing two resources and thus both are necessary for the Annotation to make sense, rather than each individually. The ordering of these resources may also be important to capture in some situations.

The explicit semantics are important for text mining and applications that reason over multiple Annotations, as well as for the client to determine how many of the resources to display to the user or in which order they should appear.

The model introduces three Multiplicity Constructs to address these requirements:

1. `oa:Choice`; where only one of the constituent resources should be used

2. `oa:Composite`; where all of the constituent resources must be used, but in any order

3. `oa:List`; where all of the constituent resources must be used in the specified order. The same resource is also an `rdf:List`, to convey the ordering.

The Multiplicity Constructs can be used to be explicit about multiple Selectors, for example to give a choice between the easy to implement, low resolution rectangle, or the preferred and more accurate, but also more complex, SVG polygon.

They can also be nested to allow, for example, a Choice of either a single Selector or a Composite of two further Selectors. This would be useful in a similar scenario to the previous one, where the target resource is a Video and the choice is between a low fidelity rectangle and time range expressed as a FragmentSelector using the Media Fragments specification, or a high fidelity SvgSelector combined with the time range.

Although this model requires additional nodes, it covers all of the use cases and does not require the interpretation of individual triples to change in various circumstances. The re-use of `rdf:List` also makes the ordering use case easier to implement.

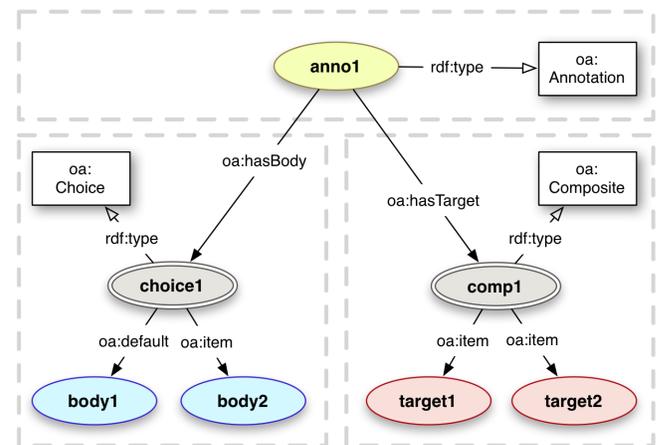

**Figure 11. Multiplicity Constructs**

Figure 11 depicts a Choice construct in the role of the Body, and a Composite in the role of the Target. Given this Annotation, a client would select either `body1` (by default) or `body2` if appropriate to display, as a comment about the set of resources comprising `target1` and `target2`, with the set's identity given as `comp1`.

## WORKED EXAMPLE

Imagine the situation where a user wants to annotate the current text of the Community Group's home page to point out that the description of the group is sadly out of date. At the current time, the description talks about milestones in the process that have already passed, as if they were in the future. In this case, the annotation is being created to request that the page be updated which is expressed with the oa:editing Motivation.

There are two parts of the description in particular that require changes. Each must therefore be identified by a Specific Resource (with UUIDs in the example) and described by a textual Selector (blank nodes :b3 and :b4). For education, we use one of each Selector. The time at which the resources are being annotated is important, as if the change were to be made, then the Annotation would no longer be valid. This is recorded using the same TimeState, the blank node :b2, for both Specific Resources. They also both have the same yellow class for CSS styling, which is described in the Style linked from the Annotation itself. Finally, as both are required to be changed, the Annotation uses the oa:Composite multiplicity construct to link the two together. The Turtle serialization for this Annotation is given in Figure 12.

This Annotation might be rendered in a manner similar to Figure 13. This is a mock-up only, however the effect would not be difficult to produce in a modern web browser.

```
<http://openannotation.org/eg/anno1> a oa:Annotation ;
  oa:isMotivatedBy oa:editing ;
  oa:styledBy <http://openannotation.org/eg/style1.css> ;
  oa:hasBody _:b1 ;
  oa:hasTarget <urn:uuid:F6C32FCA-4ED8-4B19-9716-60379E4638AE>
  .

_:b1 a dctypes:Text, cnt:ContentAsText ;
  dc:format "text/plain" ;
  cnt:chars "These two sections in yellow should be updated as this has already been done!" .

<urn:uuid:F6C32FCA-4ED8-4B19-9716-60379E4638AE> a
    oa:Composite ;
  oa:item <urn:uuid:4B356C9F-3B9C-45D1-94D7-4846E2AC65DB> ;
  oa:item <urn:uuid:7202CBC4-1AB3-44EF-956C-07E8053D6EE5> .

<urn:uuid:4B356C9F-3B9C-45D1-94D7-4846E2AC65DB> a
    oa:SpecificResource ;
  oa:hasSource <http://w3.org/community/openannotation/> ;
  oa:hasState _:b2 ;
  oa:hasSelector _:b3 ;
  oa:styleClass "yellow" .

<urn:uuid:7202CBC4-1AB3-44EF-956C-07E8053D6EE5> a
    oa:SpecificResource ;
  oa:hasSource <http://w3.org/community/openannotation/> ;
  oa:hasState _:b2 ;
  oa:hasSelector _:b4 ;
  oa:styleClass "yellow" .

_:b2 a oa:TimeState ;
  oa:when "2013-01-24T12:00:00Z" .

_:b3 a oa:TextQuoteSelector ;
  oa:exact "The effort will start by working" ;
  oa:prefix "for annotating digital resources." ;
  oa:suffix "towards a reconciliation of two proposals" .

_:b4 a oa:TextPositionSelector ;
  oa:start 488 ;      // Initially...
  oa:end 525 .        // ...proposals

<http://w3.org/community/openannotation/> a dctypes:Text ;
  dc:format "text/html" .
```

**Figure 12. Example in Turtle**

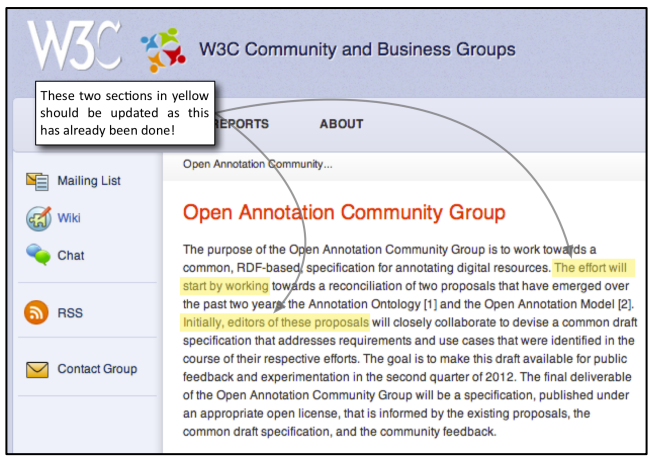

**Figure 13. Example User Interface Mock Up**

This particular Annotation is somewhat artificially complex in order to make use of as much of the data model as possible. The majority of Annotations do not require this level of description!

## CONCLUSIONS

The Open Annotation Community Group Data Model provides a single standardized and coherent method of exchanging annotations between systems. The data model scales from very simple to also accommodate more complex use cases. The principles of interoperability, such as a single way of doing something and following existing standards, provided a more objective approach to decision-making as to the design of the framework. Future work in this area is rich with possibilities from additional selectors, systems to build and process reputation models for annotators, and versioning and preservation requirements to name but a few.

**ACKNOWLEDGEMENTS**

The authors would like to thank all of the participants in the Open Annotation Community Group, and in particular the following people for their contributions to the specification: Shannon Bradshaw, Dan Brickley, Leyla Jael García Castro, Tim Clark, Timothy Cole, Phil Desenne, Anna Gerber, Antoine Isaac, Jacob Jett, Thomas Habing, Bernhard Haslhofer, Sebastian Hellmann, Jane Hunter, Randall Leeds, Andrew Magliozzi, Bob Morris, Paul Morris, Jacco van Ossenbruggen, Stian Soiland-Reyes, James Smith, and Dan Whaley.

**NAMESPACE TABLE**

| oa | http://www.w3.org/ns/oa# |
|---|---|
| cnt | http://www.w3.org/2011/content# |
| dc | http://purl.org/dc/elements/1.1/ |
| dcterms | http://purl.org/dc/terms/ |
| dctypes | http://purl.org/dc/dcmitype/ |
| rdf | http://www.w3.org/1999/02/22-rdf-syntax-ns# |
| a | http://www.w3.org/2000/10/annotation-ns# |
| ao | http://purl.org/ao/core/ |